\documentclass[preprint]{aastex}

\shorttitle{WIYN Open Cluster Study: NGC~188}

\begin{document}
\input{psfig.sty}
%%\input{epsf.sty}

\title{WIYN Open Cluster Study. XVII. \\
    Astrometry and Membership to $V=21$ in NGC~188.}


\author{Imants Platais} 
\affil{The Johns Hopkins University, Department of Physics and Astronomy,
3400 N. Charles Baltimore, MD 21218}
\email{imants@pha.jhu.edu}

\author{Vera Kozhurina-Platais}
\affil{Space Telescope Science Institute, 3700 San Martin Drive,
Baltimore, MD 21218}

\author{Robert D. Mathieu}
\affil{Astronomy Department, University of Wisconsin-Madison, 
475 North Charter Street, Madison, WI 53706}

\author{Terrence M. Girard, William F. van  Altena}
\affil{Astronomy Department, Yale University, P.O. Box 208101,
       New Haven, CT 06520-8101}


%%\altaffiltext{1}{Visiting Astronomer, Cerro Tololo Inter-American Observatory%%.
%%CTIO is operated by AURA, Inc.\ under contract to the National Science
%%Foundation.}
%%\altaffiltext{2}{Society of Fellows, Harvard University.}
%%\altaffiltext{3}{present address: Center for Astrophysics,
%%    60 Garden Street, Cambridge, MA 02138}
%%\altaffiltext{4}{Visiting Programmer, Space Telescope Science Institute}
%%\altaffiltext{5}{Patron, Alonso's Bar and Grill}


\begin{abstract}

We present techniques for obtaining precision astrometry using 
old photographic plates from assorted large-aperture reflectors in
combination with recent CCD Mosaic Imager frames. At the core of
this approach is a transformation of plate/CCD coordinates into a
previously constructed astrometric reference frame around the open
cluster NGC~188. This allows us to calibrate independently the Optical
Field Angle Distortion for all telescopes and field correctors used
in this study. Particular attention is paid to computing the
differential color refraction, which has a marked effect
in the case of NGC~188 due to the large zenith distances at which this 
cluster has been observed.
Our primary result is a new catalog of proper motions and 
positions for 7812 objects down to $V=21$ in the 0.75 deg$^2$ 
area around NGC~188. 
The precision for well-measured stars is 0.15 mas yr$^{-1}$ for proper
motions and 2 mas for positions on the system of the Tycho-2 catalog. 
In total, 1490 stars have proper-motion membership probabilities 
$P_\mu\ge10\%$.  The sum of membership probabilities
indicates that NGC~188 contains $\sim$1,050 stars down to $V=21$. 
Comprehensive lists of the candidate blue stragglers and red giant stars
substantially enlarge the number of such stars known in NGC~188.
We have also obtained a small correction
to the proper motions from the mean `motion' of background galaxies. Thus
the absolute proper motion of NGC~188 is $\mu_x^{\rm abs}=-2.56\pm0.2$ and
$\mu_y^{\rm abs}=+0.18\pm0.2$ mas yr$^{-1}$. 

\end{abstract}



\section{Introduction}

Since the seminal paper by Sandage (1962) first showed that NGC~188 
($l=122\fdg9$, $b=+22\fdg4$) is one of the oldest Galactic open clusters, 
it has been a subject of numerous 
studies and is a classical reference to the old Galactic disk population.
NGC~188 was chosen by the WIYN Open Cluster Study 
(WOCS -- Mathieu 2000) as a cornerstone cluster. To date, two WOCS 
broad-bandpass photometry papers of this cluster have appeared 
(von Hippel \& Sarajedini 1998, and Sarajedini et al. 1999).
Although NGC~188 is a fairly rich cluster and has a well-defined main sequence
in the color-magnitude diagram (CMD), this advantage starts to disappear at 
$V$$>$18 (see Sarajedini et al. 1999) where the field star contamination 
gets severe.
   
Two astrometric studies by Upgren, Mesrobian \& Kerridge
(1972) and Dinescu et al. (1996) have addressed the issue of cluster 
membership based upon the proper motions
from plates taken with the 30-in Thaw refractor of the Allegheny 
Observatory.  In order to achieve a reliable 
separation between the cluster and field stars, high-precision 
proper motions, good to 0.5 mas yr$^{-1}$ or better, are required.
This is a consequence of a very small tangential velocity
of NGC~188 relative to field stars -- only $\sim$3 mas yr$^{-1}$.
Both studies are successful in identifying the bright members of NGC~188, 
i.e., giant branch, upper main sequence, and blue stragglers.  
However, the main sequence turn-off for NGC~188 is at $V\sim15.5$, 
which is barely
brighter than the limiting magnitude of the Allegheny plates. To solve
this problem, one can use deep photographic plates taken with large
reflectors, as shown by Dinescu et al. (2000) in the case of the
globular cluster Palomar 12. However, photographic plates have been
completely phased out as a detector and the size of available single-chip 
CCDs is too small to cover efficiently the field-of-view of large telescopes.

In this study, we explore an approach by combining archival plates
with CCD mosaic images, which are now routinely available on large
telescopes. In terms of astrometry, there are several layers of difficulties
stemming from this technique. First, the astrometric properties of
CCD mosaics are not well understood yet. We have spent considerable effort
and time to calibrate the geometry of the NOAO CCD Mosaic Imager at the
Kitt Peak National Observatory Mayall 4 m telescope (Platais et al.
2002). One vexatious conclusion from this study is that the CCD mosaic
metric may not be stable and requires frequent calibration. Second, the
focal planes of nearly all large telescopes are geometrically distorted, with
radial pincushion distortion prevailing. This distortion
must be taken out prior to calculating the proper motions. Third,
NGC~188 is located $\sim$5$\degr$ from the north equatorial pole, 
resulting in a large zenith distance at all times and thus requiring
a considerable correction for atmospheric refraction, including its
color dependence. Since the cluster can be observed at the upper and lower 
culminations, the direction of the refraction vector can vary by up to 
$180\degr$. Not accounting for this would lead to color-related systematics
in positions. Finally, an uneven distribution of cluster members across
the field-of-view creates a specific problem related to the
choice of a reference frame. In purely differential calculations of
proper motions it is desirable to use tentatively selected cluster members
as such a reference frame. With few members in the outskirts of a cluster
this is not feasible but adding field stars may introduce an unwanted
bias in the proper motions as a function of position.  

In order to avoid this, we use an external astrometric reference frame
constructed earlier (Platais et al. 2002). If all sets of cartesian 
coordinates are transformed
into this frame, it is expected that all coordinates and the subsequent
proper motions will also be on the system of our reference catalog 
coordinates, i.e.
the International Celestial Reference Frame (ICRS).  Realization of
this concept is the basis of our astrometric reductions leading to
proper motions and positions down to $V=21$ in the 0.75 square degree 
(deg$^2$) area around NGC~188.  Since several aspects of data reduction are
new in the context of cluster astrometric studies, we give all necessary 
details to aid in future applications. The cluster membership is then 
calculated for all well-measured stars. By combining $BV$ photometry,
astrometric membership, and radial velocity data we provide a new 
census of the blue straggler and red giant population in NGC~188.

\section{Astrometric Reductions}

\subsection{Observational material and measurements}

 The astrometric reductions are based on 30 photographic plates taken 
with three different large reflectors in Johnson's $BV$ bandpasses,
combined with numerous CCD mosaic frames
obtained with the Kitt Peak National Observatory (KPNO) Mayall 4-m telescope 
(Table~1).  The deepest KPNO 4-m telescope plate, 3140, reaches
$B=23$ and was selected to form a master input catalog containing
over 8,000 objects, including stars and galaxies. 
This catalog covers a circular area with a radius $30\arcmin$,
centered on $\alpha=0^{\rm h}44^{\rm m}20^{\rm s}$ 
$\delta=+85\degr18\farcm9$ (equinox J2000).  The cluster center is located
$\sim$$6\arcmin$ off this center.  Thus, the total area of the 
sky covered by this study is $\sim$0.75 deg$^2$. The useful area
of the Hale 5-m telescope plates is that of a smaller
circle having a radius of $11\arcmin$. 

Recognition should be given to an earlier effort to
digitize the KPNO 4-m plates at Kitt Peak (Twarog 1983), for the sake of better
photographic photometry and their potential use in future proper-motion 
studies.  Since the Yale 2020G PDS microdensitometer is upgraded with
laser metrology, which substantially enhances the measuring precision,
we decided to re-scan all of the KPNO 4-m plates,
in addition to the Mt. Wilson 60-in and Hale 5-m telescope plates. 
All plates were scanned in a fine-raster, object-by-object mode. 
A zeropoint drift in this measuring engine is controlled and accounted
for using frequently repeated measurements of several check stars. Image
positions were determined with the Yale Image Centering routine
(Lee \& van Altena 1983), which performs a two-dimensional Gaussian 
fit to the pixel data. 

The CCD observations were obtained with the National Optical Astronomy
Observatory (NOAO) 8K$\times$8K CCD Mosaic Imager on the Mayall 4-m
telescope at KPNO (Table~1). This CCD Mosaic consists of eight thinned,
close-packed back-illuminated 2048$\times$4096 SITe CCDs, 
covering a total area of
$36\arcmin\times36\arcmin$ on the sky. Since the photographic plates 
encircle a two times larger area, a five pointing montage of CCD Mosaic 
frames has been constructed to achieve complete coverage of the entire
field. The exposure times range from 10 sec to 180 sec. 
The PSF fitting technique was chosen
to obtain positions and instrumental magnitudes on each of the 856
individual CCD chips among the total of 107 CCD Mosaic frames. The
details of PSF fitting are given in Platais et al. (2002).   

\subsection{Calibrating the sets of coordinates}

It is apparent that the mixture of observational data used in this study
is not tractable by an ordinary plate-to-plate solution because 
of large radial pincushion distortions in the focal planes of the several 
telescopes and the extreme range of hour angles at 
which NGC~188 had been observed.  The low geographic latitudes of the
telescope sites (see Table~1) imply large zenith distances even 
at the upper culmination of NGC~188. Hence, full consideration
must be given to an accurate accounting for refraction, including the
color effects. Such effects are not possible to correct without
knowing the colors of stars.  As indicated in Table~1, NGC~188 was observed 
in the broadband Johnson $BV$ and thus \bv color indices
are readily available. 

Perhaps, the most straightforward and rigorous approach for handling such
a heterogeneous collection of cartesian coordinates is to calibrate each set
of coordinates against a distortion-free astrometric flat field. In the
course of studying the NOAO CCD Mosaic Imager metrics we created the Lick
astrometric reference frame around NGC~188 (Platais et al. 2002) consisting
of 1863 stars down to V=17. The positions and proper motions of this 
reference frame are on the system of the ICRS, as close as it can be 
represented by $\sim$80 Tycho-2 stars in the vicinity of NGC~188
(H\o g et al. 2000). Thus, the coordinates from each plate and CCD Mosaic 
frame were calibrated using this astrometric standard, as described next.

\subsubsection{Differential color refraction}

Unlike other studies where the target coordinates are precorrected for
refraction, we elected to add the expected amount of differential 
refraction (relative to the field center) to the equatorial
coordinates of reference stars, prior to calculating their 
standard coordinates via a gnomonic projection. 
Such an approach makes good use of our firm knowledge of the 
equatorial coordinates and $\bv$ indices for all of the reference stars.
For the atmospheric refraction
calculation, we used a simple method described by Stone (1996). This
method requires only knowledge of the zenith distance, a spectral class
or a color index, and meteorological conditions. It should be noted
that a `gray' differential refraction, i.e., ignoring the color effects,
can be easily accounted for by the lower-order plate constants, whereas
the color-related part of refraction manifests itself as non-linear 
systematics correlated
with color. There is a danger, though, in adding an explicit color term 
to the plate constants because such a term may also be
a consequence of magnitude-dependent systematics known as `magnitude
equation' in plates or charge transfer effect (CTE) in CCD detectors.
In most frequently used $B$ and $V$ bandpasses, the differential color
refraction (DCR) between an A and M spectral type star can reach 50-100 mas
even at moderate zenith distances.  Therefore, the effect of differential 
color refraction in high-precision astrometry must be precorrected,
as has been advocated by Stone (1996, 2002). We used the code 
developed by Stone (1996) with a few modifications related to the spectral
energy distributions of stars, which is a primary source of DCR in
broadband astrometry, especially at short wavelengths. In the K-M 
spectral range we added a finer grid of empirical spectral energy 
distributions and also improved the color-to-spectral type transformation 
ensuring a smooth conversion at all colors. 

We note that in order to obtain precise absolute atmospheric 
refraction, the meteorological conditions (ambient temperature,
atmospheric pressure and dew-point readings) should be recorded at the
time of observations (Stone 1996). Sometimes partial or even all
meteorological readings can be missing, especially for old photographic
plates. In such cases, we used the mean seasonal readings for a particular
observing site.   

\subsubsection{Plate scale and mean distortion coefficients}

A high-precision astrometric reference frame allows us to derive improved 
plate scale and distortion coefficients, which are indispensable in
astrometric studies not otherwise having an external standard upon which
to rely.  The only study to address the issue of distortion coefficients for
the KPNO 4-m and Hale 5-m telescopes is that by Chiu (1976).
To find the Optical Field Angle Distortion (OFAD) characteristics,
we used the techniques described in Guo et al. (1993), which allow
us to eliminate higher order cross-terms due to the coordinate system
rotation and uncertainty in the OFAD center. The resulting coefficients
and associated errors are given in Table~2 in both the $B$ and $V$
bandpasses and separately for the $x$ and $y$ 
directions, corresponding to right ascension and declination, as indicated in 
column 2 (labeled as Set). 
A combination `$BV$' means that the coefficients are
nearly identical in both bandpasses and that the coefficients are
calculated using all plates. We could not find a bandpass dependence
in the scale, but there is a very small though discernible difference
between the plate scale in the $x$ and $y$ direction for the KPNO 4-m 
plates taken with the UBK-7 Wynne triplet corrector.  Similar differences
as a function of direction are also noticed for the cubic terms.  
The origin of these differences is not well-understood and
may not even be real. On the other hand, the cubic term for the KPNO
4-m `blue' plates is $\sim0.3\%$ larger than on the `visual' plates,
similar to the findings of Cudworth \& Rees (1991) for a nearly identical
Cerro Tololo Inter-American Observatory (CTIO) 4-m telescope.
It should be noted that the aperture of the Mt. Wilson 60-in telescope was 
always stopped down to 32 inches, therefore the cubic term may not be 
valid for the full aperture.  There is a trace of barrel
distortion for this telescope, albeit only in the $x$-direction.
In general, the agreement between our distortion coefficients and those
by Chiu (1976) is good. We believe that owing to a much better 
astrometric reference frame and a large number of common reference stars (up to
1,600) the reliability of our distortion coefficients is higher.

In calculating the equatorial coordinates for all sets of plate measurements,
we always first determined the OFAD center and only then applied the
mean cubic and fifth-order terms given in Table~2. As pointed out by
Guo et al. (1993), the distortion center of the CTIO and KPNO 4-m 
telescope plates should be known to better than $\pm1$ mm in order
to fully take out the distortion effects and thus eliminate the need
for higher-order plate constants. 

Finally, we note that the Hale
5-m telescope plates are of high astrometric quality with a useful
field-of-view up to 11$\arcmin$ in radius and do not show any signs
of the color terms reported by Dinescu et al. (2000), once the 
differential color refraction has been accounted for. The referee
pointed out that the distortion coefficients derived in this study
might be valid only for the epoch of our plates. The reason for
this caution is a noted instability of the Hale 5-m telescope
mirror support system at high zenith distances compounded by a 
considerable astigmatism.  We note that the Hale 5-m telescope 
cubic distortion term derived for the epoch 1970 and from a field 
near zenith (Chiu 1976,1980) shows excellent agreement with our 
study while the fifth-order term in Chiu (1976) is by $\sim$20\% 
larger than our value but so is the fifth-order term for the
KPNO 4-m telescope. Apparently more studies are warranted to
ascertain the stability of distortion coefficients for the
Hale 5-m telescope.

\subsubsection{Improving CCD Mosaic chip constants}

A prerequisite for reliable proper motions is the ability to combine
the photographic plates and recent CCD Mosaic frames.
In Platais et al. (2002) we describe in detail how to determine the
so-called CCD chip constants (position of the chip center and a rotation
angle around this center). Since this study, two modifications have
been added. First, all 26 CCD Mosaic frames in $B$ were also 
incorporated.  Astrometrically these frames are
comparable to the $V$ frames and, therefore, were used to better
determine the chip constants. The cubic distortion term in the $B$ bandpass
is ($-7.02\pm0.03$)$\times10^{-16}$ rad pixel$^{-3}$, or 
($-2.46\pm0.01$)$\times10^{-6}$ mm$^{-2}$, which is $\sim$0.4\% larger
than in the $V$ bandpass (Platais et al. 2002).  There is no perceptible
difference in the value of fifth-order distortion in these two bandpasses
\footnote{In the system of pixel coordinates introduced in Platais
et al. (2002) the mean OFAD center is actually located at $x$=4190 and 
$y$=4093 pixels}. 

Second, the residual map (Fig.~5 in Platais et al.
2002) was constructed anew, restricted only to $\sim$30 CCD Mosaic frames
taken with a 15-30 s exposure time and thus minimizing the atmosphere-induced
pattern in residuals. To avoid an unrelated contribution from adjacent
CCD chips to the global pixel coordinates, all residuals were strictly 
subdivided among the eight chips and only then averaged.

\subsubsection{Calculating mean positions}

To calculate the equatorial coordinates for each plate or CCD Mosaic frame,
we followed the prescription given in \S 2.2.1 by adding the amount of 
differential refraction to each reference star and temporarily working
in a `refracted' coordinate system.
Once the equatorial coordinates are obtained for all program stars,
the appropriate amount of atmospheric refraction should be, of course,
subtracted.  This requires a knowledge of \bv color indices which are 
obtained from $BV$ photographic photometry using the KPNO 4-m telescope plates 
and photoelectric $UBV$ photometry (Sandage 1962).
In the case of CCD Mosaic frames, finding the color index is more
complicated. The output of the PSF fit (see \S 2.1) consists of 
pixel coordinates and instrumental magnitudes in a given bandpass in some 
random order. To match these entries with our master input catalog, 
first, we calculate equatorial coordinates without a differential color
refraction term for all 107 CCD Mosaic frames. Then, from all sets of 
equatorial coordinates, a declination-sorted master CCD catalog is created.
A $0\farcs5$ tolerance is used to find multiple detections and, thus,
in the vast majority of cases only one entry per object is kept. The master
CCD catalog then allows us to i) identify with the master input catalog;
ii) find the corresponding $BV$ magnitudes from the existing photometry
source (\S 3.1) and calculate photometric zero-points for all eight CCD 
Mosaic chips; iii) calibrate instrumental
magnitudes and \bv color indices for each object in the master 
CCD catalog.

In the astrometric reductions (i.e. coordinate transformation) we used 
a linear plate model plus quadratic plate tilt terms, $p$ and $q$. 
The higher-order distortion was pre-corrected prior to the plate solution. 
Restriction to the low-order-only
plate constants in combination with a large number of 
reference stars ($\sim$1,600 for plates and $\sim$950 for CCD Mosaic 
frames), ensures a reliable transfer of the ICRS to much fainter
magnitudes over the entire field-of-view. This approach does not
require an a priori knowledge of possible cluster members, which
then, at least in theory, simplifies the maintenance of proper-motion 
zeropoint in the periphery of the cluster. This is a well-known problem,
if one has to rely upon the pre-selected cluster stars as a reference
frame. A centrally concentrated spatial distribution of cluster stars 
necessitates the inclusion of field stars in the outskirts of the cluster 
into the reference frame 
and, hence, may introduce a bias in proper motions as 
a function of position. The effect of such a variation in the 
proper-motion zeropoint is increasingly harmful as the precision of
proper motions rises, since the intrinsic dispersion of field-star proper
motions cannot be reduced.

\subsection{Proper Motions and Cluster Membership}

In total, there are 137 sets of equatorial coordinates in the ICRS system at 
the epochs of the plates and CCD Mosaic frames. Once all detections for
a given star are found, the proper motion is simply the slope from
a linear fit to positions as a function of time (epoch) in years. Since
the quality of an object's image can change from set to set, weights
have been assigned to each individual image following the scheme 
outlined in Platais et al. (1998). In our case, after each fit 
the residuals are accumulated as a function of height (peak value) of 
a two-dimensional Gaussian fit to the
digitized photographic images, or as a function of instrumental magnitude 
in the case of CCD Mosaic frames. The so-called `scatter curves' (\S 3.1 in
Platais et al. 1998) for each plate or CCD frame are the sources for
the weight estimates. We choose a 4$\sigma$ threshold to eliminate
obvious outliers, always starting with the worst one. An elimination
rate exceeding 1-2 datapoints for a star usually is a sign of disturbance
by an adjacent image or is an indication of a saturated image. Such cases 
are discussed in more detail in \S 4.1.

\subsubsection{Correction for magnitude equation}

Once the first-cut proper motions were calculated, it was evident
that many photographic plates showed a considerable magnitude equation
in positions, indicated by the dependence of residuals on the Gaussian peak
density value (Fig.~1a,b). A fiducial curve was drawn to remove this 
dependency, as shown by Kozhurina-Platais et al. (1995),
and the proper-motion solution repeated using the corrected positions.
We note that stars of all magnitudes might suffer from magnitude 
equation since only the mean of all reference-star residuals (see
\S 2.2.4) has been forced to be zero.  Nearly all plates are found
to be affected by a magnitude equation of different degree and sign
in declination but only $\sim$30\%
of the plates show a weak magnitude equation in right ascension. This
is counterintuitive to the notion that a magnitude equation is a
guiding-error induced effect, although slightly elongated images 
are common on many of our plates. 
We emphasize that the magnitude equation cannot be eliminated entirely 
in the way described here.  In effect, we mostly reduce the scatter
due to magnitude equation and hence improve the precision but not 
necessarily the accuracy of the positions. 

It is likely that a mixture of plates taken at various hour angles 
and with three different telescopes would generate some resultant 
magnitude equation in proper motions.  What really matters is not 
so much its 
absolute value as the ratio of `magnitude equation to standard error 
of proper motions'. If this ratio is high, say, over 3, 
the proper motions of cluster members affected by such a magnitude 
equation could be misplaced in the vector-point diagram so much
that it would yield a very low or even zero membership probability
due to the intrinsic rarity of over-$3\sigma$ events.

Given the manner in which the star positions have been
calculated, it is the faint stars which are most prone 
to this effect. To examine this, we used the photometric members of NGC~188
selected from the preliminary $BV$ photometry described in \S 3.1.
The proper motions of 
these stars clearly indicated the presence of a magnitude equation up 
to $\sim$0.7 mas mag$^{-1}$ (Fig.~2). The proper motions were, thus, corrected
for the magnitude equation relative to the stars with $V=16$ --
the approximate mean magnitude of the reference stars.

\subsubsection{Spatial bias in proper motions}

An analysis of preliminary proper motions for the 
photometrically-selected members,
confronted with their radial-velocity membership (see \S 3.2), quite 
unexpectedly revealed the existence of a considerable spatial bias in 
the proper motions.
In other words, a number of bona fide cluster members located near the 
edges of the field had proper motions incompatible with their 
radial-velocity membership status.
A closer look at this bias indicated mostly a linear dependence of
proper motions on both coordinates, $xy$, reaching up to 2 mas yr$^{-1}$.
It is likely that the origin of this bias is related to the 
Tycho-2 proper motions, which were used to update the positions of 
reference stars in the reductions of the Lick plates. If around NGC~188 
there is a spatial bias in the proper motions of Tycho-2 stars, it 
will propagate through the Lick reference frame stars right into the 
final proper motions.  One way to
take out this bias is to correct the proper motions themselves. Since
throughout reductions, including the construction of our standard frame
from the Lick plates, 
we always used a linear plate model plus quadratic plate-tilt terms, 
it is expected that any spatial bias in the proper motions
should also comply to the same model.  To derive the amount
of spatial bias, we used the proper motions of photometric cluster 
members (supplemented with the radial-velocity membership when 
possible) and solved for the linear and plate-tilt terms under 
an assumption that the proper motions of cluster stars must be the 
same everywhere in the field.
Figure~3 shows that after correction by these terms, there is no
gross trend in the proper motions of 342 radial-velocity probable cluster 
members ($V<16.5$). It should be mentioned that some systematic spatial 
pattern at the level of a few tenths of a mas yr$^{-1}$ is still present 
in our proper motions.  We could not pinpoint the source of this pattern 
even if, for instance, we eliminate specific sets of plates when
calculating the proper motions. Note that a 0.1 mas yr$^{-1}$ systematic
in proper motions will produce an error in tangential velocity equal 
to $\sim$0.8 km s$^{-1}$.  Given the expected three-dimensional velocity 
dispersion for an old open cluster on the order of 1 km s$^{-1}$ 
(cf. Girard et al. 1989), we caution that our proper motions  
should not be used to obtain a reliable estimate of the internal velocity 
dispersion.

\subsubsection{Precision of proper motions and positions}

The formal error of each proper motion is simply the standard error
in the slope of a linear fit to the weighted positions. The distribution
of proper motion errors is presented in Fig.~4a. 
The highest precision of our proper motions is 0.15 mas yr$^{-1}$  
per coordinate. The lower envelope of the error distribution represents 
the well-measured and isolated stars near the center of cluster, which
is densely covered by the plate material and CCD Mosaic frames.
In the outer parts of the cluster there is no coverage by the Hale 5-m
telescope plates and also the number of CCD frames is low; therefore
the errors are higher, as indicated by a band-like structure in the
error distribution (Fig.~4a). Although we have measured over 7,800
objects, only those with proper-motion errors less than 
1.2 mas yr$^{-1}$ are considered in the subsequent membership analysis. 
This appears to be the precision limit above which only some definite field
stars can be identified by their large proper motion but no reasonable 
detection of cluster stars is possible, since membership probabilities 
for such stars never exceed $\sim$20\% (see \S 2.3.5.). 

The precision of positions is dependent upon the chosen epoch and is
optimal at the mean epoch of all positions used in the linear fit.
The best positional precision in either coordinate at the mean
epoch is 2 mas (Fig.~4b). It should be noted that the mean epoch for
the inner area ($\sim36\arcmin\times36\arcmin$; see also \S 2.1) is
1995.5, whereas outside this square the mean epoch drops to 1986.0,
which is a consequence of a lower total number of CCD frames used to derive
the mean positions.  We do not know the 
contribution by uncertainties in the Lick reference frame positions, 
therefore it is only provisionally estimated that our positions are within
$\sim5-10$ mas on the system of the ICRS in the area of NGC~188. 
For practical applications the mean epoch and the associated formal
positional error is not needed since even the largest error is merely
100 mas. However, for the aficionados of precision astrometry we provide
the necessary data to interpolate the errors to any chosen epoch. 

\subsubsection{Proper motions of galaxies}

The depth of this proper motion survey allows us to identify numerous 
faint external galaxies. Nominally, our proper motions are absolute since 
all positions are reduced into the inertial ICRS system presumably
well-represented by the Tycho-2 stars. 
The proper motions of galaxies provide an excellent test
of this assumption. In total, we are able to visually identify 99
galaxies, excluding the spirals and irregular ones which are not suitable
for astrometry. Among the selected galaxies, 70 have formal proper-motion
errors ranging between 0.5 to 1.7 mas yr$^{-1}$.  All are fainter than
$V\simeq17$.  The color range of these galaxies is 0.5$<$\bv$<$1.8. It was
a surprise to find that the proper motions of galaxies have a magnitude
equation in declination -- approximately equal to the applied 
magnitude equation correction in the stars (see Fig.~2)! 
On the other hand, it has been long suspected but
not proven (Girard et al. 1998, Platais et al. 1998) that stars and
galaxies may obey different magnitude equations.  Our data
suggest that the galaxies are nearly free of a magnitude equation but
the stars are not.  Hence we undid the magnitude correction
which had been applied to galaxies.  No less surprising is the mean 
`proper motion' of galaxies itself, equal to $-2.75\pm0.22$ mas yr$^{-1}$ 
in $x$ and $-0.59\pm0.18$ mas yr$^{-1}$ in $y$.  Since
we cannot expect to detect the motion of galaxies, this is just a
measure of how much the system of mean positions and proper motions of 
Tycho-2 stars in the area around NGC~188 differs from the ICRS.

\subsubsection{Membership probabilities}

The vector-point diagram (VPD) for the proper motions derived in this study is
shown in Fig.~5.  Our combination of the high precision of the proper
motions and a careful accounting for the systematic errors leads to
the compact appearance of the cluster stars in the VPD.
From a variety of membership probability estimators (e.g.,
Kozhurina-Platais et al. 1995, Dinescu et al. 1996 and references
therein) we have chosen to use a local sample concept that employs
two-dimensional Gaussian frequency functions for cluster and field,
(Kozhurina-Platais et al. 1995).  The central idea of the local-sample
approach is, for each target star, one selects
the stars which share the properties of that target star, e.g.,
the brightness and spatial location, and use only this representative
sample in calculating the membership probability. In this way, it is
possible to remove a well-known bias in membership probabilities due
to the differences in the shape of the cluster and field star luminosity
functions.  In the study by Kozhurina-Platais et al. (1995) a moving
bin in magnitude and spatial position was used to select the local
sample, the bin being centered on the target star.
Effectively, in this case, the most important parameter in
membership calculation -- the proper-motion error -- alone can be used
to select the local sample.  This is made possible by
the following two correlations. First, there is a clear correlation
in the proper-motion error distribution (Fig.~4a), showing larger
errors at fainter magnitudes. Thus, proper-motion error 
can used to form the equivalent
of a magnitude bin. Second, the proper-motion error also has
a spatial dependence (Fig.~6). Toward the field edges the proper
motion-error gradually increases and, hence, this error can be used
to account for a spatial distribution of cluster members.
Finally, we note that selecting a sample of proper motions with
similar errors solves a complicated statistical problem, if a large
range of measuring errors is simultaneously considered (cf.
Dinescu et al. 1996).

A critical part of any membership
calculation is an accurate representation of the field stars. Usually
the parameters of the field-star distribution are calculated
simultaneously with that of the cluster stars.
However, the mean motion and proper-motion dispersion of the
field stars can be estimated apart from the membership probability
formalism. We have used
the color-magnitude diagram to isolate field stars and then found the
center and dispersion of their proper-motion distribution at a range of
magnitudes.  It appears that the distribution of field star proper motions
may require an $+8\degr$ rotation around the mean center to align
it with the $xy$ axes and eliminate the cross-terms in a Gaussian
representing them. Nonetheless, it was not applied due to an 
uncertainty of this angle of comparable size.

In the case of cluster stars, we take advantage of the fact that the
center of their proper-motion distribution is constant at all magnitudes.
As with the field stars, we also
estimated the dispersion of photometrically-selected cluster members stars,
but this time in 0.2 mas yr$^{-1}$ wide error bins.
The magnitude-dependence of the dispersion is smoothed and parametrized for
the subsequent membership probability calculation.
The adopted cluster center in
the vector-point diagram is $\mu_{x}^{c}=-5.31$ and
$\mu_{y}^{c}=-0.41$ mas yr$^{-1}$. Only stars with proper-motion
errors less than 1.2 mas yr$^{-1}$ are considered in the membership
probability calculation. This restriction is motivated by the small
separation between the centers of the field, $\Phi_f$, 
and cluster-star, $\Phi_c$,
distributions ($\Delta_\mu$=3.0 mas yr$^{-1}$)
and a similarly small dispersion for the bulk of the field stars
($\leq$4 mas yr$^{-1}$).  At larger proper-motion errors the two
distributions become almost indistinguishable, which is
examplified by the study of Upgren et al. (1972).
As pointed out by Dinescu et al. (1996), many highly
probable cluster members in Upgren et al. are field stars.

As a result, the only two parameters we need to determine
are the heights of the two Gaussian distributions, $\Phi_c$ and
$\Phi_f$, in the vector-point diagram (see Eq. 4 in Kozhurina-Platais
et al. 1995). This is easily done for each target star's
local sample, containing stars with nearly identical proper-motion precision.
Moving from one target star to another to form a local sample, we used
a varying bin size in error ranging from 0.15 to 0.3 mas yr$^{-1}$,
such that the bin size increases for lower-precision proper-motion samples.
Once the local sample is selected, we bin its vector-point diagram with
a bin size equal to $0.8\varepsilon$, where $\varepsilon$ is the
dispersion of the cluster proper motions having a proper-motion error very
close to that of the target star.  It is the number of datapoints in each 
VPD bin what is fitted to the sum of two Gaussians mentioned above.
The resulting membership probability is defined as

\begin{equation}
P_\mu=\frac{\Phi_c}{\Phi_c+\Phi_f}.
\end{equation}

A good indicator of the cluster and field separation is the maximum 
membership probability as a function of magnitude (Fig.~7). 
As seen in this plot, very high membership probabilities ($P_\mu > 90\%$)
extend down to $V\sim19$ , i.e.  at least three magnitudes deeper than 
prior studies.  At fainter magnitudes, the maximum probability gradually
declines and at the limiting $V=21$ is only $\sim30\%$ -- in conformity 
with the local sample conception.  To isolate cluster members at such 
faint magnitudes, proper-motion membership should be coupled with the 
color-magnitude diagram of the same stars.  A formal sum of membership 
probabilities indicates that NGC~188 contains $\sim$1050 stars down 
to $V=21$.  

\subsubsection{Cluster membership completeness}

Due to the imposed 1.2 mas yr$^{-1}$ limit on the precision of 
proper motions, cluster membership has been calculated for only $\sim$70\% 
of all stars in our catalog. How does that affect the cluster
membership completeness?  The vast majority of stars having 
proper-motion errors higher than this limit are fainter than $V=20$.
However the fraction of such stars is spatially dependent.
This is illustrated by the fraction of stars, $\phi$, not considered in the
proper motion membership calculation, as a function of magnitude (Fig.~7).
If we examine the stars within the inner circle (radius=$13\arcmin$)
around the cluster center (i.e., the area covered by the Hale 5-m telescope
plates), our membership study begins to become incomplete at $V=19.5$;
fainter than this the completeness decreases rapidly.
On the other hand, outside this inner circle where the proper-motion
precision is generally poorer, incompleteness sets in at $V\sim17$ but
does not become severe until $V=18.5$.  In general, the spatial
dependence of this completeness is more complicated since the majority of CCD
Mosaic frames, taken with the KPNO 4-m telescope, are on the same central field 
center (see \S\S 2.1, 2.3.3) and hence the membership completeness also 
drops beyond the $36\arcmin \times 36\arcmin$ FOV.

Our membership calculation procedure also excludes a few bright 
cluster members with poor measurements and, hence, a large proper motion 
error.  All such cases are due to some serious problem, e.g., saturated
or overlapping images, therefore alternative means such as photometry 
and radial velocities should also be invoked to decide on membership.

The final result is a catalog\footnote{The catalog is available
on the WWW at \texttt{http://www.astro.yale.edu/astrom/}.} (also Table~3) 
of positions and proper motions of 7812 stars and galaxies.
This smaller number of objects in our catalog as compared to the master 
input list (see \S 2.1) is entirely due to the initial inclusion of 
very bright stars (and their secondary images from the Racine wedge) 
and extremely faint objects not present on the other plates.
Membership probability is given only if the mean proper motion
error does not exceed 1.2 mas yr$^{-1}$. Complete cross-identifications
are provided with the Dinescu et al. (1996) catalog and the frequently
used list of stars from Sandage et al. (1962). 

\section{Non-astrometric Data}

\subsection{Photometry}

Although photometry is not the main subject of this study, we encountered 
issues for whose resolution the magnitudes and colors are indispensable. 
In other words, the high-precision 
ground-based astrometry cannot achieve its potential without precise
photometry.  Among our CCD Mosaic frames, there are 26 frames in $B$
and 81 in $V$.  At the time of observations it was not possible to
obtain photometric standards, except for the one night of October 10, 2001.
The spatial extent and density of stars within each of the photometric 
standard fields (Landolt 1992) are not well-suited to calibrate the Mosaic's 
individual CCD chips. 
To calibrate our photometry, we used the $BV$ CCD photometry obtained
by Hainline et al. (2000) at the KPNO 0.9-m telescope.  This study
covers a $40\arcmin\times40\arcmin$ area centered on
NGC~188. The offset between this photometry and our instrumental
magnitudes, for each individual CCD chip,
provides the zeropoint correction. The additional color
terms needed to put our photometry onto Johnson's standard $BV$
system, were provided by P. Massey (private communication). It is
estimated that our CCD $BV$ photometry is precise to 0.01 mag but 
less so for $V<13$ due to increasing saturation of images, and for the
faintest stars ($V>20$).  Full analysis of our $BV$ photometry will
be presented elsewhere.

To illustrate the quality of the cluster membership determination, 
we provide several color-magnitude diagrams (CMD). The color-magnitude
diagram for all objects (stars and galaxies), having a reliable
determination of $\bv$ in our catalog, is provided in Fig.~8. 
The main sequence NGC~188 is very well visible down to $V\sim20$;
however, at fainter magnitudes it is hidden among the numerous 
Galactic disk dwarfs. Then Fig.~9 shows the CMD for 1490 stars
stars with membership probabilities $P_\mu\ge10$\%.  
A rather low cutoff in probabability is chosen because of our desire 
to preserve the low-probability faint cluster members, which would 
be lost applying the traditional cutoff at $P_\mu =50\%$ (see Fig. 7). 
This CMD shows a clean main sequence down to $V=21$ and distinct
populations of subgiants, red giants, and blue stragglers.
The hot blue subdwarf at $V=16.3$ is confirmed to be a cluster member.
Finally, the color-magnitude diagram of 
very likely field stars ($P<10\%$) does not show any trace of
the cluster's main sequence (Fig.~10).
It should be noted that traces were seen in our earlier data
(Platais et al. 2000), which prompted us to reconsider the astrometric 
reductions and develop techniques decsribed here.
A detailed analysis of the CMD will be presented elsewhere (Platais et al.,
in preparation).

\subsection{Radial Velocities}

Since 1996 precise radial velocities for stars brighter than $V=16.5$ have
been obtained with the WIYN 3.5-m telescope at Kitt Peak, Arizona and the Hydra
multi-object spectrograph. This instrumentation provides a resolution 
of 20,000 at
wavelengths around the Mg B triplet (Mathieu 2000). More details
about the instrumental setup and reduction technique can be found
in Meibom et al. (2001).
The corresponding rms velocity precision for a representative single 
cluster member is 0.45 km s$^{-1}$ (Meibom et al. 2000). 
The uncertainty in the WIYN radial velocity zeropoint relative to the 
IAU radial-velocity standards is found to be within $\sim$0.3 km s$^{-1}$. 
To date over 5,000 observations have been collected for more than 500 
stars in the area of NGC~188. Among these, we have identified 342 members
with radial velocities clustered  around the (preliminary) mean velocity 
of $-42.4$ km s$^{-1}$.  Owing to this large negative along-the-sight
velocity and its high precision, the radial velocities are essential for 
identifying true cluster members among the blue stragglers and 
the red giant branch which in the case of NGC~188 has a rather 
complex appearance.  

\section{Discussion}

\subsection{Comparison with Dinescu et al. (1996)}

The study by Dinescu et al. (1996), entirely superseding the work by
Upgren, Mesrobian, \& Kerridge (1972), is presently the source 
of the highest precision proper motions in NGC~188. 
There are 1082 stars common with
this study. To examine both catalogs in terms of membership
probabilities, we have limited the number of common stars to 413 -- those
having comparable precision of proper motions ($\sigma_\mu<$1 mas yr$^{-1}$)
in the two catalogs. In general, agreement between the two
estimates of membership probabilities, $P_\mu$, is remarkably good.
Thus, about 83\% of selected common stars show differences in
membership probabilities smaller than  $\Delta P_\mu=25\%$.

Nonetheless, it is important to understand the reasons for large differences
in membership probabilities, i.e., when $\Delta P_\mu>50\%$. 
There are 39 such stars and 35 of them have radial velocity measurements, 
that provide conclusive evidence of their membership. 
What can we learn from these stars?  First of all, there is a set of 
seven stars for which our proper motion data are
not reliable. These include very close pairs of images. On photographic
plates our image centering algorithm has a limited ability to ignore 
an adjacent, partially overlapping image. Therefore, the calculated
center frequently refers to the mean center of both images or is biased
to some degree. On CCD Mosaic frames this is much less of a concern if 
the PSF fitting approach is used, as in our case.  Of course, if on 
the early-epoch photographic plates the measured position of an image is 
significantly distorted, then having a correct position on the CCD 
Mosaic frames does not help to obtain reliable proper motions.
Among all 7812 catalog stars we have identified
41 cases whose proper motions are hopelessly wrong and have replaced them
with $\mu_x=0.0$ and $\mu_y=0.0$. Some of such stars, having an apparent
single image, are located in the vignetted part of photographic plates.
It is suspected that many more 
proper motions may suffer a similar bias albeit at a much smaller scale, 
therefore all stars having a nearby image within $\sim$5$\arcsec$ are 
marked in Table~3 and should be considered with caution. We note that
in the catalog of Dinescu et al., several such stars are not resolved 
and thus their mean proper motion and calculated membership probability 
are perhaps more realistic.

Second, among the remaining 32 stars (from our sample of 39 stars with
discordant probabilities) having no problems with image crowding it 
is not obvious why the membership probabilities are in disagreement. 
The radial-velocity membership indicates that in $\sim$70\% of cases
our $P_\mu$ values correctly reflect their membership status,
whereas, for the same stars, the Dinescu et al. catalog predictions
appear to be successful at a rate of $\sim$30\%. For stars brighter
than $V$$\approx$15, which is about one magnitude brighter than the faint 
limit of Dinescu et al. study, we recommend considering both catalogs on equal
footing; i.e., if either catalog indicates cluster membership for
a star, it should be checked by alternative means such as radial 
velocities.

\subsection{Membership of variables, blue stragglers, and red giants}

NGC~188 is well-known for an unusually high number of W~UMa-type variable 
stars (Zhang et al. 2002). Among the 19 variables listed by Zhang et al.
(2002) and Kaluzny (1990), we have measured all stars except the 
variable V19, which is located outside our field.  
Table~4 contains membership probabilities of 18 variables 
(with prefix "V" from Zhang et al. 2002); 11 of them are confirmed cluster 
members.  NGC~188 contains at least six W~UMa-type
variables and one detached eclipsing binary. Very recently
Kafka \& Honeycutt (2003) published a list of newly found variable
stars in the region of NGC~188. In Table~4 these variables
are listed with a prefix "WV". A large fraction of them are genuine 
cluster members.

Sandage (1962) was first to point out the presence of blue stars above the
main sequence -- the blue stragglers.  Eggen \& Sandage (1969) identified 
11 blue straggler candidates, whereas Dinescu et al. (1996) expanded that
list to 15 stars. Here we have attempted to select all possible blue straggler
candidates, based on the following criteria:

\begin{enumerate}
\item Color bluer than $\bv=0.64-0.68$ so that the main sequence is
   entirely excluded. The stars located {\it under} the main sequence but not 
   fainter than $V=16.5$ are also considered.
\item Proper-motion membership probability higher than $P_\mu$=50\%.
\item Radial velocity indicates cluster membership.  The star is considered 
   a cluster member, if its mean radial velocity (or gamma-velocity in the case 
   of spectroscopic binaries) is within 2.5 km s$^{-1}$ from the mean 
   cluster velocity (see \S 3.2).
\end{enumerate}

A few stars with membership probabilities lower than $P_\mu$=50\%  are also 
included as blue straggler candidates, 
if their radial velocities are consistent with cluster membership.
Such stars could be among our astrometric problem stars (\S 4.1)
or located near the field edges.  For stars near the main sequence
turnoff there is an ambiguity in their identification as blue
stragglers or photometric binaries. As opposed to Eggen \& Sandage 
(1969), we have opted not to include such stars (e.g., 4506, 5080, 5101)
among our blue straggler candidates.  In total, 35 stars satisfying all three
of these conditions, or at least the first two when a radial 
velocity measurement is not available, are given in Table~5. In this table, 
we give our number (ID), cross-identication with Dinescu et al. (1996) 
catalog (N$_{\rm D96}$), astrometric membership probability ($P_\mu$), and
radial velocity membership status, RV$_{\rm m}$.  We note again
that radial velocity data are very reliable in identifying blue 
stragglers, although in the cases of binaries extended monitoring 
is usually required to firmly establish their membership.
The number of bona fide blue straggler members of NGC~188 
(i.e., with $P_\mu>90\%$ and similarly high membership
probability from Dinescu et al. (1996) and established radial-velocity
membership) is at least 19.  Thus NGC~188 offers a rich sample of blue 
stragglers, comparable to that in the Galactic globular clusters 
(Ferraro, Fusi Pecci, \& Bellazini 1995).

For years the study by McClure (1974) has served as a canonical source 
list of red giants in NGC~188, encompassing nine established cluster members.  
The actual number of red giants is much larger as indicated by the 
Dinescu et al. (1996) estimate of about 40 red giants. In this study
we provide a new census of red giants (Table~6) comprising a total
of 55 such stars. A red giant is included in this list if it is brighter
than $V=14.65$, redder than $\bv=0.95$, and is a cluster member
according to our proper motions and/or radial velocities. The
chosen color and magnitude limits are very close to those in
Dinescu et al. (1996) study.  For a few stars missing our radial 
velocity data, such information was provided by E. Green and
J. Sperauskas (private communications). 
Similarly to Table~5 we also included a few low membership probability
stars with the same motivation as for the blue stragglers. 
Faint red giants ($V>14.7$) are not included in Table~6 because of 
a possible confusion with subgiants.
The brightest red giant in NGC~188 is star 6175 at $V=10.8$.

\subsection{Absolute proper motion of NGC~188}

Following the discussion of the galaxy `proper motions' in \S 2.3.4, we 
can obtain a new measurement of the absolute proper motion
of NGC~188. Thus, subtracting the mean proper motion of the galaxies
from the mean motion of the cluster $\mu_{x}^{c}$, $\mu_{y}^{c}$, its
absolute proper motion is  $\mu_{x}^{\rm abs}=-2.56\pm0.2$ 
and $\mu_{y}^{\rm abs}=+0.18\pm0.2$ mas yr$^{-1}$. 

Baumgardt, Dettbarn, \& Wielen (2000) have provided first direct
estimate of the absolute proper motion of NGC~188 using two Hipparcos
stars. Only one of them, HIP 4349 (6175 in our catalog),
is a bona fide cluster member; however, its large proper motion error 
($\sigma_\mu\sim1.7$) in the Hipparcos catalog (ESA, 1997) prevents us from 
using this star in direct comparison.  Our absolute proper motion
differs considerably (by $\Delta\mu=4.6$ mas yr$^{-1}$) from the 
value derived indirectly from the stellar kinematics by 
Upgren et al. (1972). 

In light
of the  considerable magnitude-equation correction necessary for the
faint stars but not the galaxies (see \S 2.3 and Fig.~2), it can be
questioned how well we actually determine the absolute proper motion of
NGC~188.  Unable to compare directly Hipparcos and our proper 
motions, we use the 
Lick reference frame (Platais et al. 2002) for external checks. 
A total of five Hipparcos stars yield the following differences
in proper motions in the sense `Hipparcos -- Lick': 
$\Delta\mu_x=+0.80\pm0.37$, $\Delta\mu_y=+0.13\pm0.74$ mas yr$^{-1}$. These
differences indicate that the system of Lick proper motions is
reasonably close to the ICRS and a small correction in declination, 
deduced from the galaxy proper motions, is consistent with Hipparcos data.
In right ascension such consistency is less evident, although it is 
this coordinate, in which the correction in faint-star proper motions
is relatively small (see Fig.~2).  Thus we estimate that the absolute
proper motion of NGC~188 is probably accurate only to 0.5-1 mas yr$^{-1}$. 

\subsection{Center of cluster}

Although seemingly easy to derive, a precise estimate of the cluster 
center location cannot be found in the literature. Two subtle problems
make the determination of a cluster center non-trivial. First, our 
field-of-view is not centered on the cluster (see \S 2.1), so that
some undetected peripheral cluster members may reside beyond its near 
edge.  Second, the spatial distribution of probable cluster members 
is slightly asymmetric. 

Two techniques were used in deriving the cluster
center. The first consists of the mean center coordinates
of a series of concentric samples of cluster members, the center of each
sample defined as the median position along right ascension and 
declination. For this purpose, we formed two initial samples, selecting 
stars near the main
sequence (Fig.~9) having membership probabilities $P_\mu\ge10\%$,
and then selecting those with
$P_\mu\ge75\%$, supplemented with the blue stragglers and red giants
from Tables~5,6. These two samples were trimmed at a number
of radii ranging from $23\arcmin$ to $15\arcmin$, using an initial estimate
of the cluster center.  The median position of each trimmed subsample was
calculated and the mean of all these median measures determined.
The process was iterated to convergence. A more rigorous approach is 
to fit the cluster's apparent density profile in both
directions. We assume a simple Gaussian profile that gives us
a reasonable approximation of the observed density profile, which
is true only to the extent that the asymmetry in the real profile
can be ignored.
Nonetheless, both approaches yield a consistent cluster center,
equal to $\alpha=0^{\rm h}47^{\rm m}12\fs5$ and
$\delta=+85\degr14\arcmin49\arcsec$ (J2000) with the estimated standard 
error from the scatter of derived cluster centers
indicated in Table~7. This is $41\arcsec$ away 
from the center of Ring I (Sandage 1962), usually adopted as the center
of NGC~188. The basic astrometric data are summarized in Table~7.

\section{Conclusions}

This study serves a two-fold purpose. First, we have shown that
precision astrometry is feasible using a combination of various
old photographic plates and recent CCD Mosaic frames. Once
properly calibrated, the CCD mosaic imagers are excellent devices
for astrometry. 

Second, we have obtained precise proper motions and new astrometric 
membership probabilities down to $V=21$ in the 0.75 deg$^2$ area around 
the old open cluster NGC~188. The total number of cluster members
down to this limiting magnitude is about 1,050. The unprecedented 
depth of accurate proper motions now allows us to probe the
main sequence of NGC~188 down to $\sim0.5 M_\sun$ without any 
ad hoc assumptions about the field-star contamination.

The key to achieving a high accuracy in this case
of heterogeneous observations of NGC~188, all obtained at high zenith
distances, is a careful correction for differential color 
refraction. This requires a knowledge of star colors, which 
is to say that photometry at least in two bandpasses is essential
to astrometry in order to fully realize its potential and to avoid
a possible color-related systematic bias. In addition, a prior
knowledge of the so-called photometric cluster members greatly
facilitates the detection and correction for systematic errors such as 
a magnitude equation. 

Using a diverse set of first-epoch photographic material leads to an
array of significant systematic errors, which made this study far
more complex than expected. The chosen path of deriving proper motions 
via the astrometric reference frame is not  
infallible due to the required high accuracy of such a standard.
However, the traditional approach of using the `best plate' as an
initial reference frame is equally prone to systematic errors and
at this point is not feasible for the CCD Mosaic frames, which themselves
require a careful calibration. 

We provide a new determination of the absolute proper motion of NGC~188, 
calibrated against background galaxies. 
Our proper motions support the membership of six W~UMa-type eclipsing
binaries found in NGC~188, thus confirming the high rate
of these variables in this open cluster. NGC~188 also contains at least
19 blue stragglers, which undoubtedly will provide new insights in the
origin of these unusual stars.

\acknowledgments

We thank Allan Sandage for loan of the Hale 5-m and Mt. Wilson
60-in telescope plates. We appreciate the use of the KPNO 4-m
telescope plates which were kindly forwarded to us by Bruce Twarog.
We are grateful to Betsy Green and Julius Sperauskas for providing 
us radial-velocity data prior to their publication.  
We gratefully acknowledge support by the
National Science Foundation under grants AST-0321794 (I.P.) and
AST-9812735 (R.D.M.). We thank the referee, Kyle Cudworth, for his
insightful comments and suggestions which have helped to improve the paper.
This research made use of the SIMBAD database operated at CDS, 
Strasbourg, France, and also of the Base Des Amas  (WEBDA) 
developed and maintained by J.-C. Mermilliod.



%%\appendix

%%\section{Appendicial material}

\begin{figure}
\epsscale{0.8}
\plotone{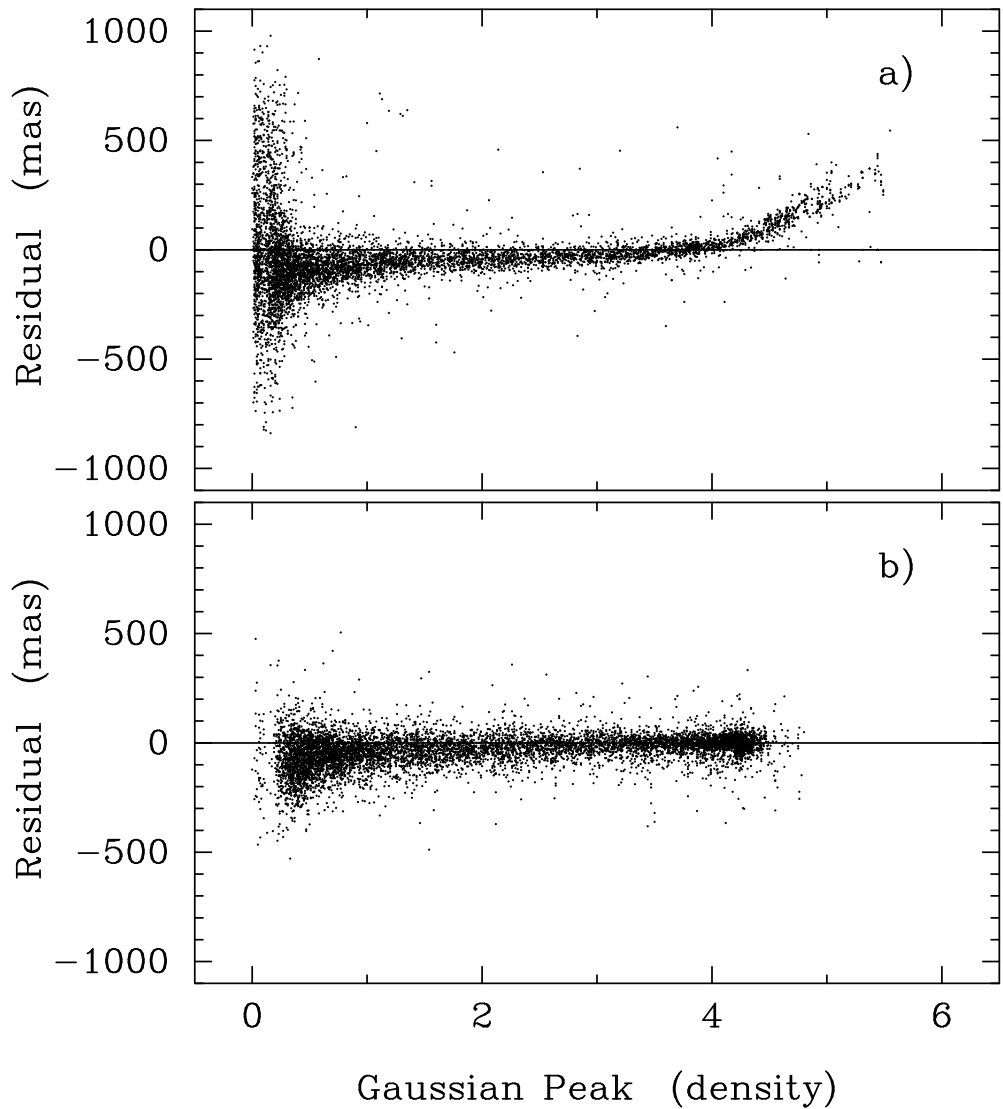}
\caption{Coordinate residuals in declination after the preliminary proper 
motion solution: a) between the KPNO 4-m telescope plate 1626 and mean
coordinates at the epoch of the plate; b) the same but for the KPNO plate 3140.
These two plots exemplify an extreme and a weak magnitude equation in
the coordinates. In each case, a fiducial curve is drawn to remove the
coordinate dependence as a function of magnitude.}
\end{figure}

\begin{figure}
\epsscale{0.8}
\plotone{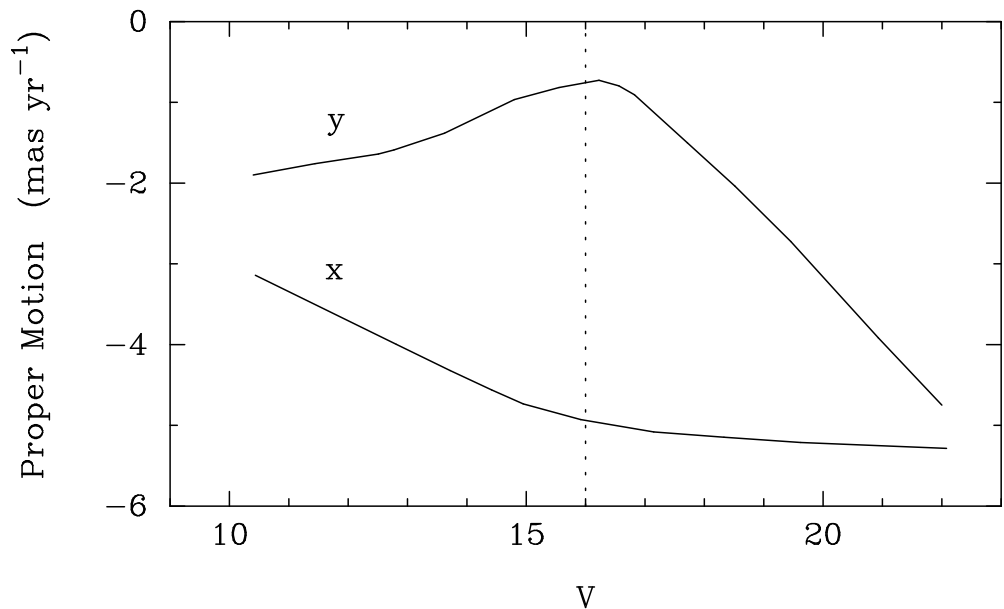}
\caption{Mean magnitude equation in proper motions derived from the
photometrically-selected preliminary cluster members
in right ascension ($x$-direction) and declination ($y$-direction).
Note a large magnitude equation in the declination proper motions for
faint stars. The dotted line marks the mean magnitude of reference
stars, relative to which the proper motions were corrected for
magnitude equation.}
\end{figure}

\begin{figure}
\epsscale{0.8}
\plotone{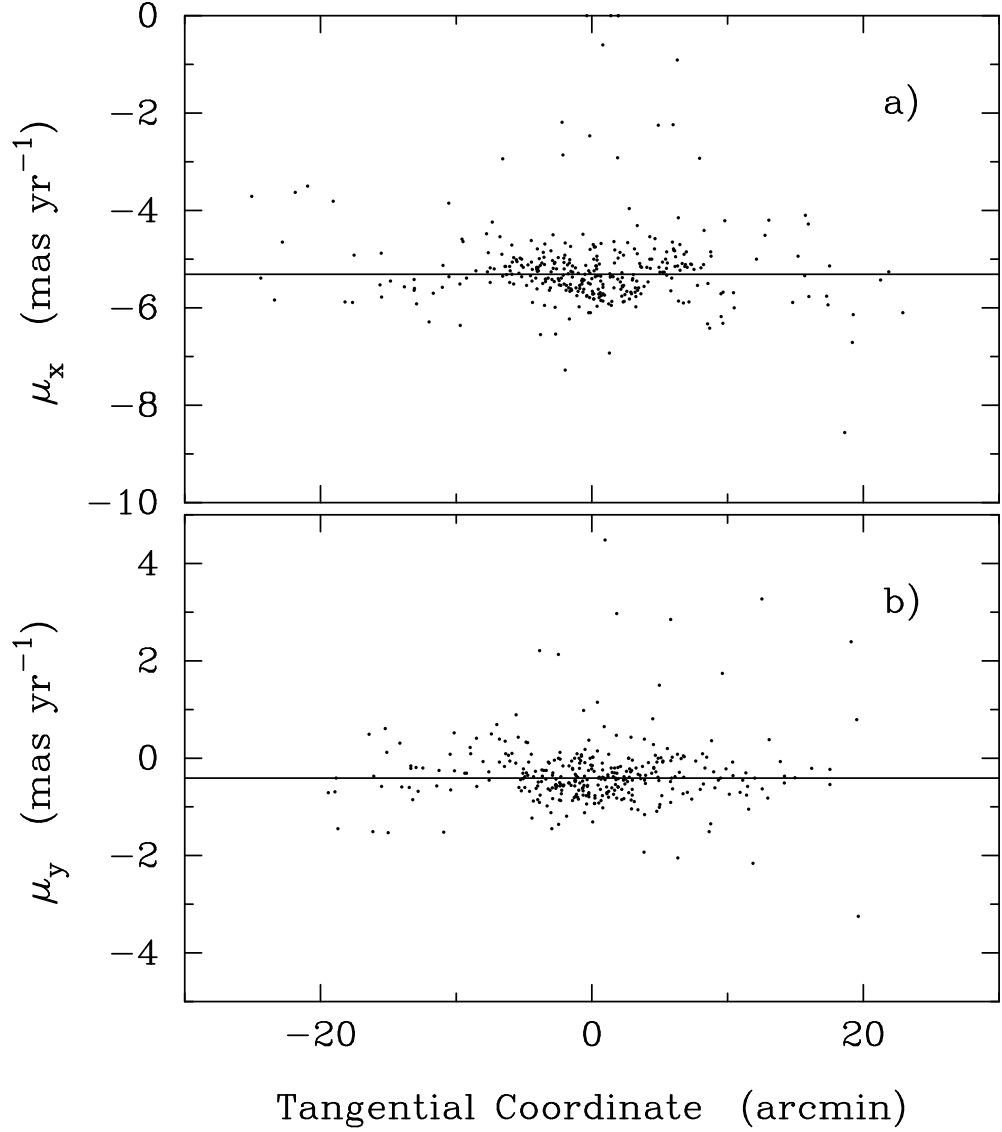}
\caption{Proper motions as a function of tangential coordinates for
342 radial-velocity cluster members: a) in right ascension 
($x$-direction); b) in declination ($y$-direction).
There still seems to be present a non-random
pattern in the distribution of proper motions, perhaps indicating some
small residual systematic error. The horizontal line shows the
adopted cluster mean proper motion.}
\end{figure}

\begin{figure}
\epsscale{0.8}
\plotone{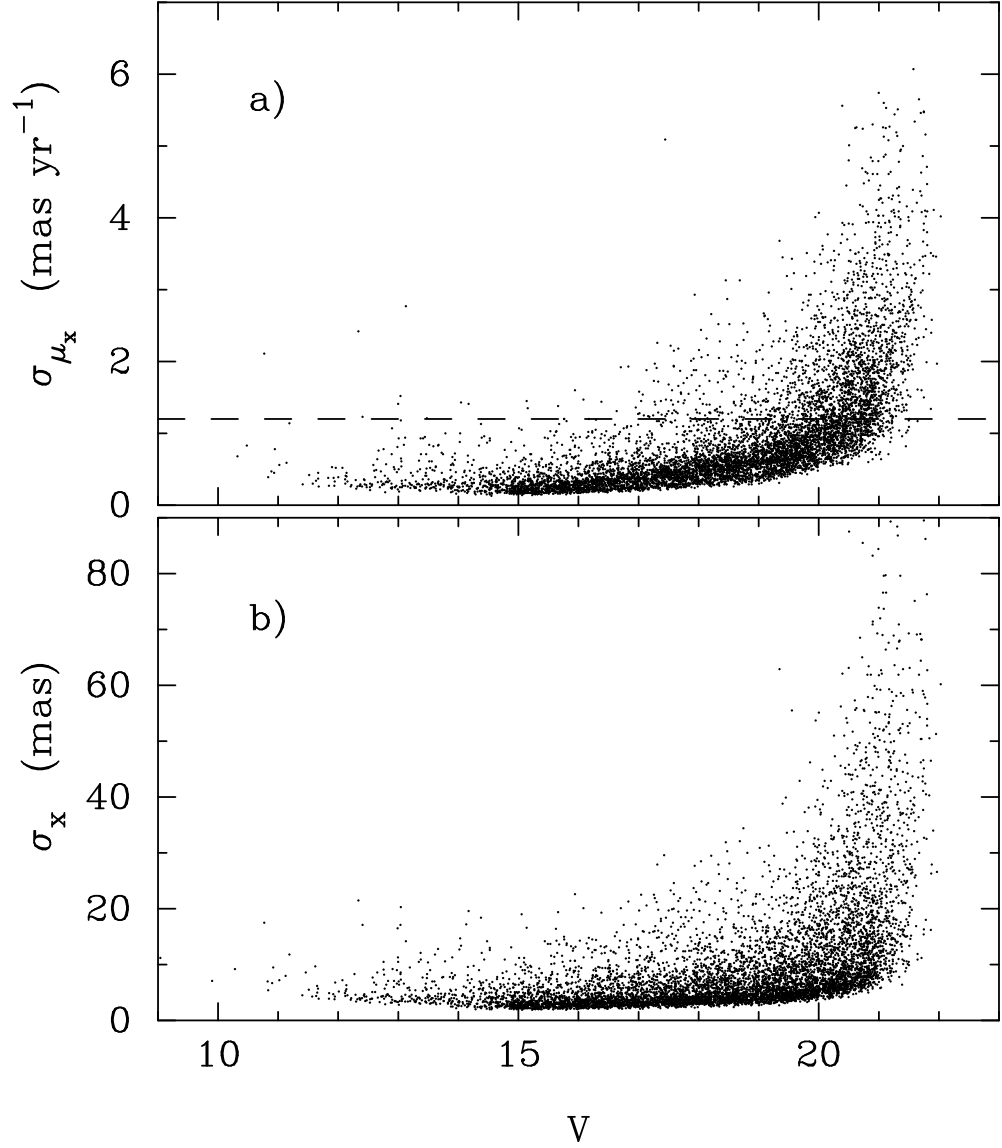}
\caption{Standard error distribution in right ascension as a function 
of $V$-magnitude: a) in proper motions. Cluster membership
probabilities are calculated only for stars with $\sigma_{\mu}<1.2$
mas yr$^{-1}$, as marked by the dashed line; 
b) positional error distribution.}
\end{figure}

\begin{figure}
\epsscale{0.8}
\plotone{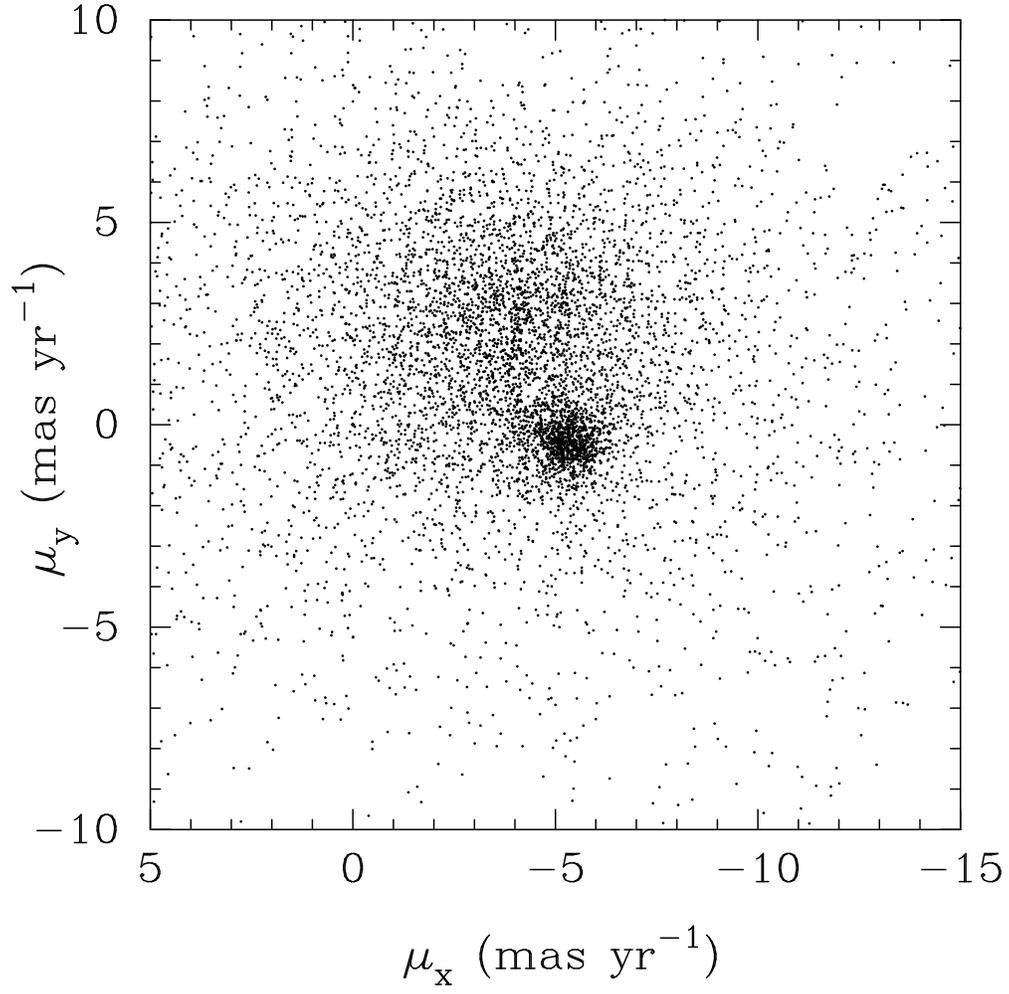}
\caption{Proper-motion vector point diagram (VPD) in the area of NGC~188.
A tight clump at $\mu_x=-5.3$ and $\mu_y=-0.4$ mas yr$^{-1}$ indicates 
the cluster location in the VPD. The center of field stars is located 
only 3.0 mas yr$^{-1}$ away from the cluster centroid.
}
\end{figure}

\begin{figure}
\epsscale{0.8}
\plotone{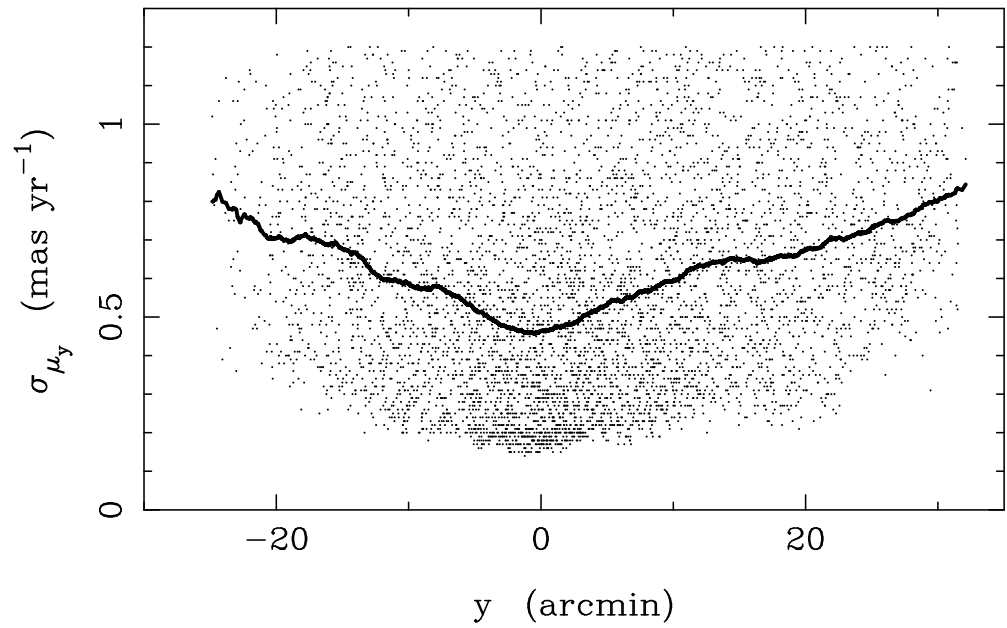}
\caption{Proper-motion error distribution as a function of declination
($y$-coordinate). The dark curve is showing the average error calculated
in a moving $5\arcmin$-wide bin. The smallest errors are at the cluster
center ($y$=0.0).
}
\end{figure}

\begin{figure}
\epsscale{0.8}
\plotone{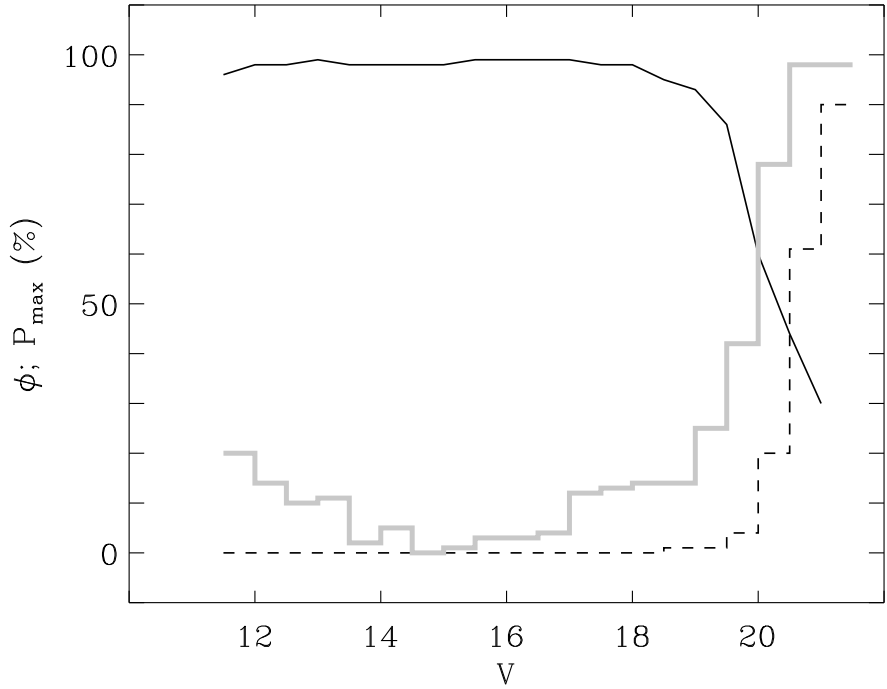}
\caption{Maximum membership probability, $P_{\rm max}$ (in per cent),
as a function of magnitude (continuous curve). 
At $V=19$ and fainter the maximum probability is gradually getting 
lower, as expected due to the growing contamination by field stars. 
Also shown is the fraction of stars in 20 magnitude intervals, 
$\phi$ (in per cent), with proper motion errors exceeding 
1.2 mas y$^{-1}$ within a radius of $r=13\arcmin$ from the cluster
center (dashed histogram) and outside this radius (shaded histogram).
These histograms show where our membership study becomes incomplete
with fainter magnitude and increasing distance from the cluster center.
}
\end{figure}

\begin{figure}
\epsscale{0.8}
\plotone{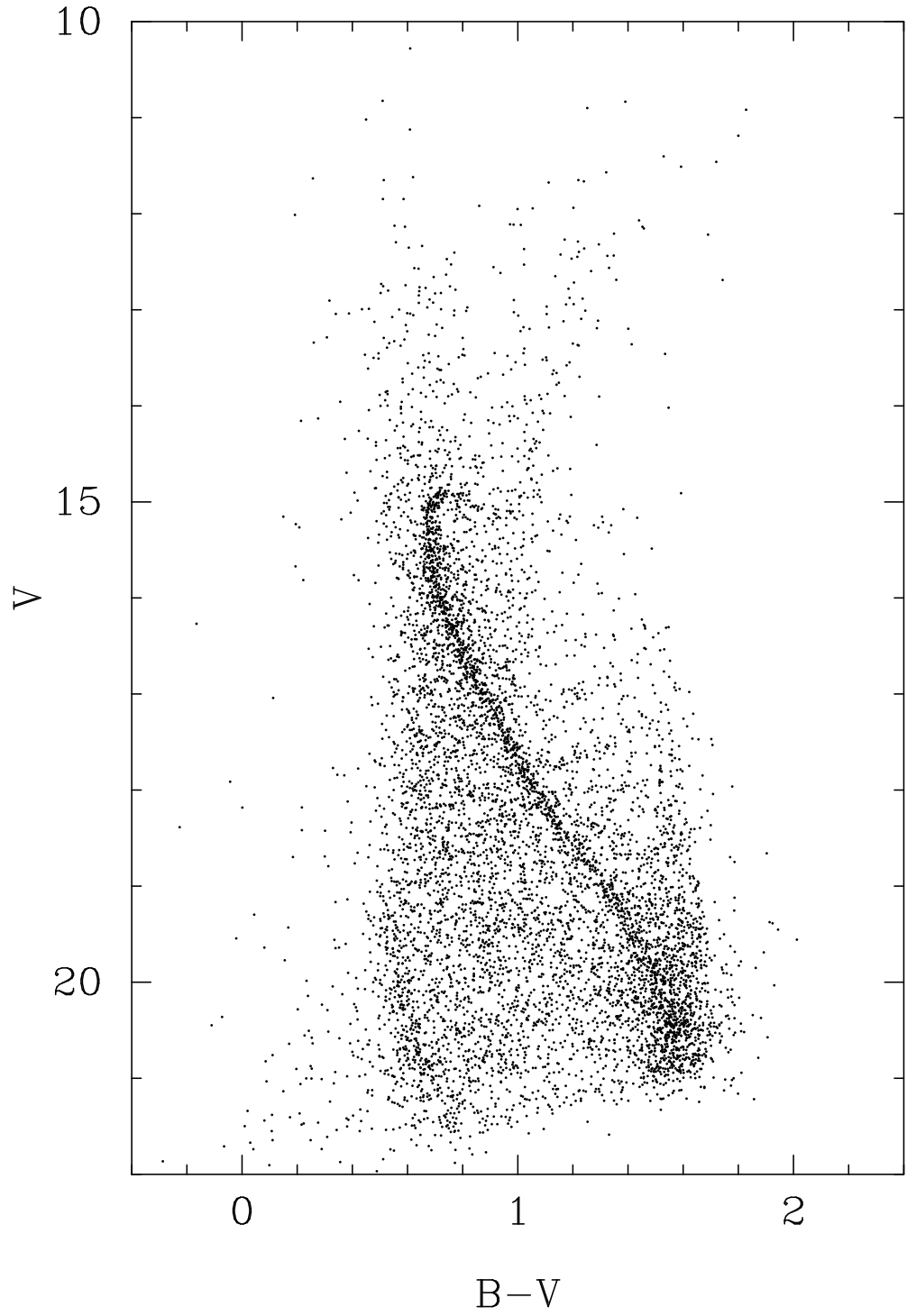}
\caption{Color-magnitude diagram for all stars (and galaxies)
in the catalog with reliable \bv color indices. Only main sequence
is clearly visible down to $V=20$.
} 
\end{figure}

\begin{figure}
\epsscale{0.8}
\plotone{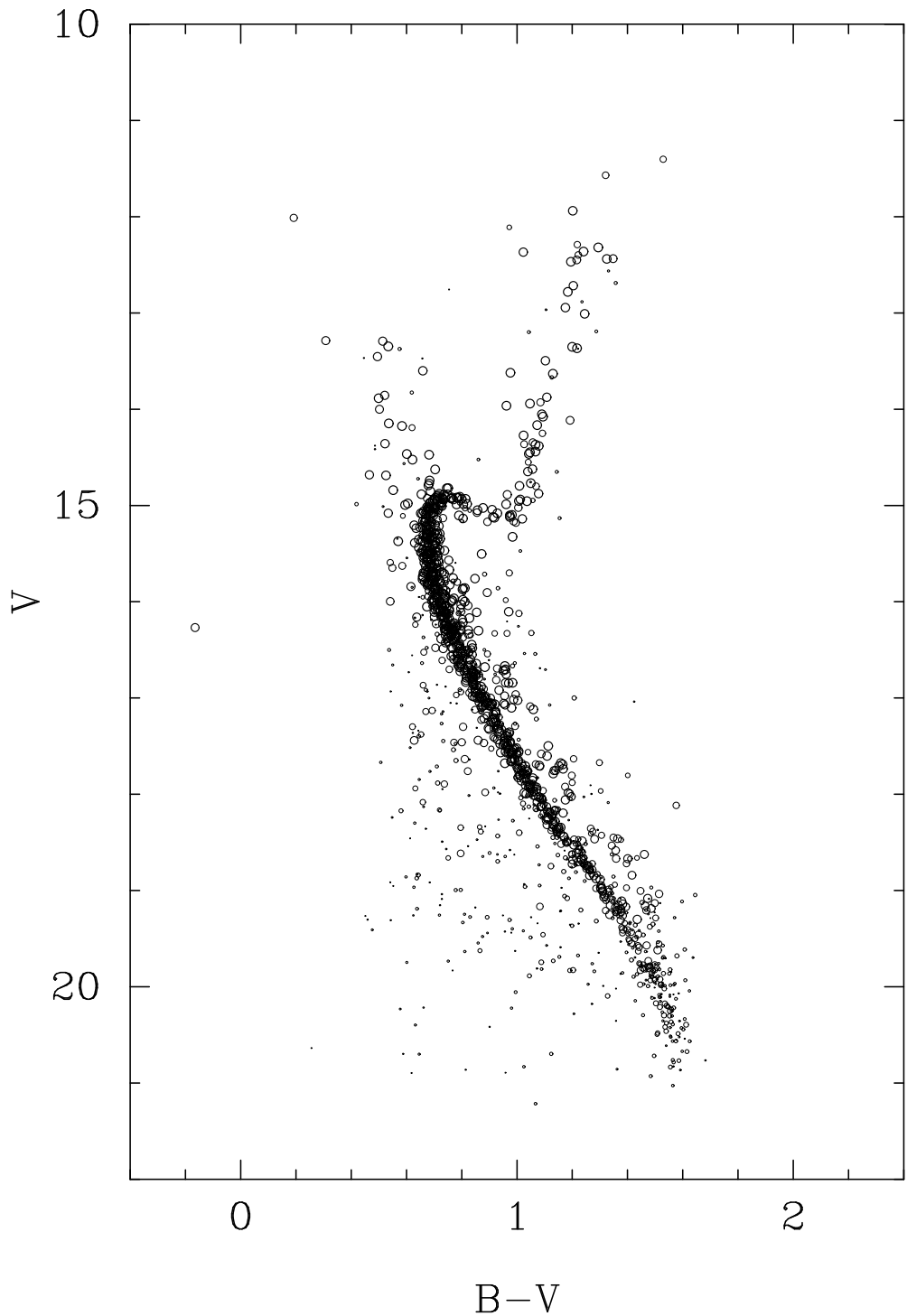}
\caption{Color-magnitude diagram for 1490 probable cluster
members with $10\%\le P_{\mu} \le99$\%. The size of each circle is 
proportional to the membership probability.  Note the presence of 
numerous blue stragglers and the high membership probability of 
a hot subdwarf (star 4918 = Sandage II-91) on the blue side of the CMD.} 
\end{figure}

\begin{figure}
\epsscale{0.8}
\plotone{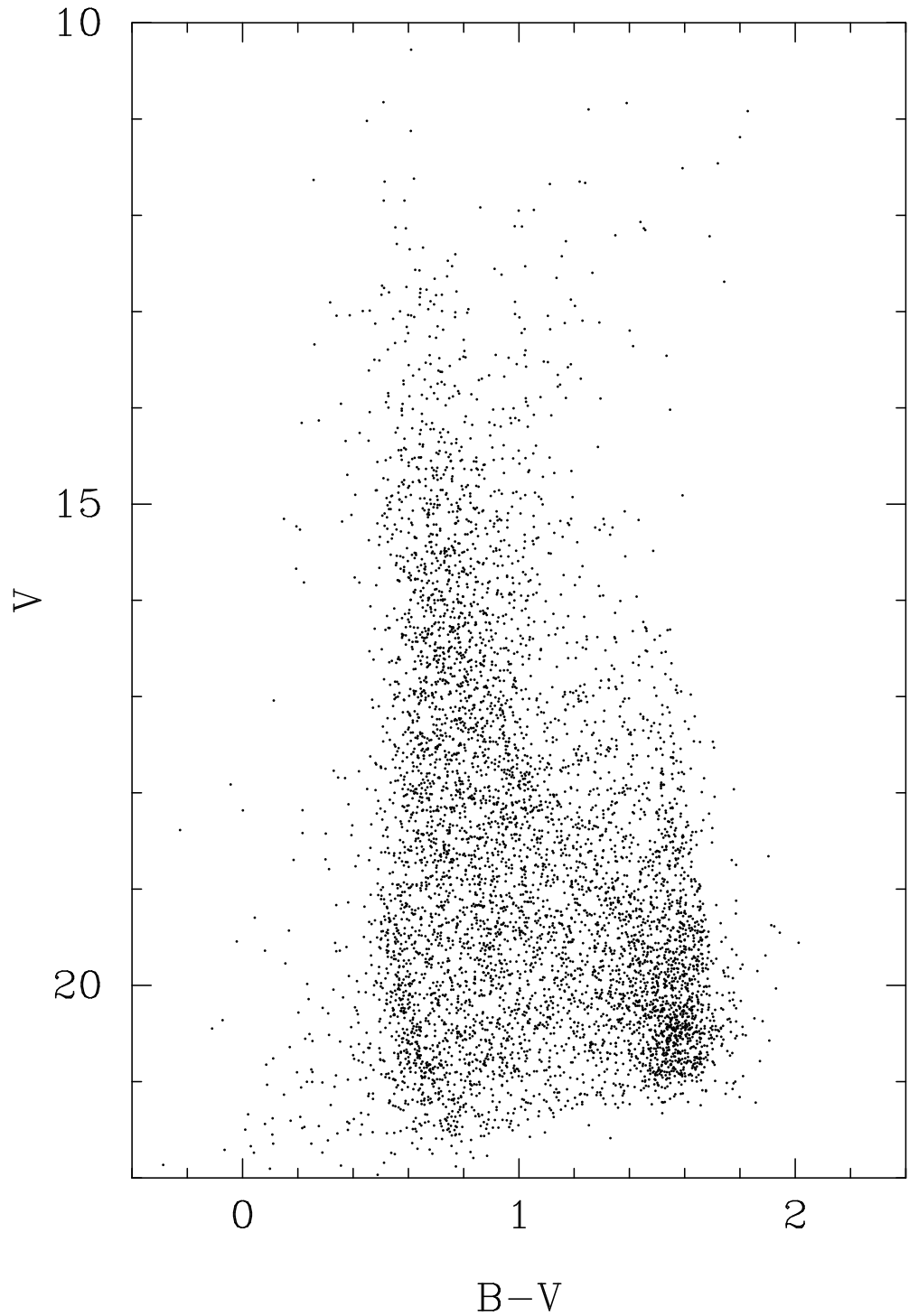}
\caption{Color-magnitude diagram of likely field stars with
$P_\mu<10$\% and the stars with $\sigma_{\mu}>1.2$ mas yr$^{-1}$.
The lack of any trace of the cluster's main sequence is a sign of 
success in the attempt to separate cluster stars from the field.
}
\end{figure}

\clearpage
\begin{deluxetable}{lcccc}
\tablewidth{4.8in}
\tablenum{1}
\pagestyle{empty}
\tablecaption{Plates and CCD Mosaic frames}
\tablehead{
\colhead{Telescope}& \colhead{Latitude}  & \colhead{Plates} & 
\colhead{Epoch Range} & \colhead{Scale} 
  }
\startdata
Mt. Wilson 60-in & $34\fdg22$  & 5$B$,2$V$ & 1958-1959 & $27\farcs1$ mm$^{-1}$\\
Hale 5-m & 33.36 & 2$B$,2$V$ & 1960 & $11\farcs1$ mm$^{-1}$ \\
KPNO 4-m & 31.96 & 10$B$,9$V$ & 1975-1979 & $18\farcs6$ mm$^{-1}$\\
KPNO 4-m & 31.96 & 26$B$,81$V$ & 1998-2001 & $0\farcs26$ pix$^{-1}$\\
\enddata
\end{deluxetable}

\clearpage
\begin{deluxetable}{lccrc}
\tablewidth{6.4in}
\tablenum{2}
\pagestyle{empty}
\tablecaption{Distortion Coefficients}
\tablehead{
\colhead{Telescope}  & \colhead{Set} & \colhead{Scale ($\arcsec$/mm)} & 
\colhead{Cubic Term (mm$^{-2}$)} & \colhead{Fifth Order (mm$^{-4}$)}  
  }
\startdata
Mt. Wilson 60-in & $x/B$ & $27.120\pm0.011$  & 
($+1.3\pm1.4$)$\times10^{-8}$ & \nodata \\
& $x/V$ & \nodata & ($+1.8\pm1.0$)$\times10^{-8}$ & \nodata \\
Hale 5-m & $x/BV$ & $11.128\pm0.007$  & 
($-6.98\pm0.11$)$\times10^{-6}$ & ($-1.29\pm0.11$)$\times10^{-10}$ \\
& $y/BV$ & \nodata & 
($-6.93\pm0.09$)$\times10^{-6}$ & ($-1.55\pm0.15$)$\times10^{-10}$ \\
KPNO 4-m & $x/B$ & $18.598\pm0.002$ & 
($-3.673\pm$0.017)$\times10^{-6}$ & ($-2.23\pm0.10$)$\times10^{-11}$ \\
& $y/B$ & \nodata & ($-3.656\pm$0.017)$\times10^{-6}$ & 
($-2.21\pm0.12$)$\times10^{-11}$ \\
& $x/V$ & $18.596\pm0.002$ & 
($-3.662\pm$0.018)$\times10^{-6}$  & ($-2.35\pm0.06$)$\times10^{-11}$ \\
& $y/V$ & \nodata& ($-3.646\pm$0.014)$\times10^{-6}$  & ($-2.33\pm0.11$)$\times10^{-11}$ \\
\enddata
\end{deluxetable}

\clearpage
\begin{deluxetable}{crcccrc}
\tablewidth{3.5in}
\tablenum{4}
\pagestyle{empty}
\tablecaption{Membership of variables}
\tablehead{
\colhead{Var} & \colhead{ID}  & \colhead{$P\mu$ (\%)} && \colhead{Var} &
\colhead{ID} & \colhead{$P_\mu$ (\%)} 
  }
\startdata
V1 & 4873 &  98 && WV9  & 5481 & \phantom{0}0 \\
V2 & 4982 &  82 && WV10 & 4329 & 86 \\
V3 & 5361 &  \phantom{0}0 && WV11 & 5361 & \phantom{0}0 \\
V4 & 5337 &  96 && WV12 & 4997 & 85 \\
V5 & 4989 &  95 && WV15 & 6118 &  \phantom{0}0 \\
V6 & 5004 &  71 && WV16 & 8077 &  \phantom{0}0 \\
V7 & 5912 &  \phantom{0}0 && WV17 & 5810 & 83 \\ 
V8 & 5629 &  \phantom{0}0 && WV18 & 4751 & 97 \\
V9 & 4736 &  97 && WV19 & 6208 &  \phantom{0}0  \\
V10 & 5459 &  94 && WV20 & 5781 & 98 \\
V11 & 4705 &  98 && WV21 & 5354 & 94 \\
V12 & 5762 &  97 && WV22 & 5178 &  \phantom{0}0 \\
V13 & 5209 &  98 && WV23 & 4086 & 90 \\
V14 & 8495 &  \phantom{0}0 && WV24 & 4656 & 98 \\ 
V15 & 6571 &  \phantom{0}0 && WV26 & 3953 & 97 \\ 
V16 & 6574 &  \phantom{0}0 && WV27 & 4247 &  \phantom{0}0 \\ 
V17 & 324  &  \phantom{0}0 && WV28 & 4508 & 98 \\ 
V18 & 6842 &  \phantom{0}0 && WV29 & 4679 &  \phantom{0}0 \\ 
WV2 & 5569 &  98 && WV30 & 5800 & \phantom{0}0 \\
WV3 & 5379 &  98 && WV31 & 7290 & 52 \\ 
WV4 & 4304 &  97 && WV32 & 5516 & 24 \\
WV5 & 4742 &  98 && WV33 & 5769 & \phantom{0}0 \\  
WV6 & 4748 &  98 &&WV34 & 5767 & 98 \\  
WV7 & 4396 &  98 &&WV35 & 5876 &  \phantom{0}0 \\ 
WV8 & 4750 &  96 && 

\enddata
\end{deluxetable}

\clearpage
\begin{deluxetable}{rrcccrrcc}
\tablewidth{4.6in}
\tablenum{5}
\pagestyle{empty}
\tablecaption{Membership of candidate blue stragglers}
\tablehead{
\colhead{ID} & \colhead{N$_{\rm D96}$} & \colhead{$P\mu$ (\%)} &
\colhead{RV$_{\rm m}$} && \colhead{ID} &
\colhead{N$_{\rm D96}$} & \colhead{$P_\mu$ (\%)} & \colhead{RV$_{\rm m}$}   
  }
\startdata
451 & 1874 & \phantom{0}7 & \rm{m}\phantom{:} && 5078 & 687     &  98 &  \rm{m}\phantom{:} \\
1066 & \nodata &  68 & \nodata &&5237 & \nodata & 69 & \nodata \\ 
1888 & 1793    &  98 &  \rm{m}\phantom{:} &&5325 & 1370    & 98 &  \rm{m}\phantom{:}  \\ 
1988 & 1760    &  96 & \nodata &&5350 & 758  & 93 &  \rm{m}\phantom{:}  \\ 

2679 & 1666    &  21 & \rm{m}\phantom{:}  &&5379 & 772  & 98 &  \rm{m}\phantom{:}  \\ 
4230 & 1342    &  93 &  \rm{m}: &&5397 & 777  & 96 &  \rm{m}\phantom{:}  \\ 

4290 & 485     &  98 &  \rm{m}\phantom{:} &&5434 & 711  & 97 &  \rm{m}\phantom{:}  \\ 

4306 & 483     &  99 &  \rm{m}\phantom{:} &&5671 & 444  & 95 &  \rm{m}\phantom{:}  \\ 

4348 & 632     &  93 &  \rm{m}\phantom{:} &&5885 & 702 & \nodata & \rm{m}\phantom{:} \\ 

4447 & 1468    &  90 &  \rm{m}: &&5934 & 681 & \nodata & \rm{m}\phantom{:}\\ 

4474 & 1466    &  98 &  \rm{m}\phantom{:} && 6069 & 1321    & 79 & \nodata \\
4535 & 1436 & \nodata & \rm{m}\phantom{:} && 6882 & \nodata & 54 & \nodata \\
 
4540 & 1437    &  92 &  \rm{m}\phantom{:} &&7782 & 198     & 11 &  \rm{m}\phantom{:}  \\ 

4581 & 1346    &  98 &  \rm{m}\phantom{:} &&8104 & 237     & 91 & \nodata \\ 

4589 & 1387    &  98 &  \rm{m}\phantom{:} &&8222 & \nodata & 61 & \nodata \\

4805 & \nodata &  92 & \nodata && 8386 & 543     & 96 &  \rm{m}\phantom{:} \\
4970 & 746     &  98 &  \rm{m}\phantom{:} && 8899 & 56      & 29 &  \rm{m}\phantom{:}   \\
5020 & 721     &  90 &  \rm{m}\phantom{:} && &   & &  \\
\enddata
\end{deluxetable}

\clearpage
\begin{deluxetable}{rrcccrrcc}
\tablewidth{4.6in}
\tablenum{6}
\pagestyle{empty}
\tablecaption{Membership of red giant branch stars}
\tablehead{
\colhead{ID} & \colhead{N$_{\rm D96}$} & \colhead{$P\mu$ (\%)} &
\colhead{RV$_{\rm m}$} && \colhead{ID} &
\colhead{N$_{\rm D96}$} & \colhead{$P_\mu$ (\%)} & \colhead{RV$_{\rm m}$}   
  }
\startdata
736 & 1612    &  39 &  \rm{m}  &&5085 & 686 & 98 & \rm{m} \\ 
1141 & 1819    &  88 &  \rm{m} &&5133 & 566 & 98 & \rm{m} \\ 

1216 & \nodata &  32 &\nodata  &&5179 & 537 & 67 & \rm{m} \\ 

2442 & \nodata &  13 &\nodata  &&5356 & 763 & 98 & \rm{m} \\ 

3015 & 1161    &  24 &  \rm{m} &&5373 & 771 & 95 & \rm{m} \\ 

3037 & 1049    &  \phantom{0}1 & \rm{m}  &&5438 & 710 & 98 & \rm{m} \\ 

3046 & 1053    &  \phantom{0}0 &  \rm{m} &&5597 &1267 & 26 & \rm{m} \\ 

3062 & 1607    &  20 &  \rm{m} &&5700 & 458 & 98 & \rm{m} \\ 

3118 & 1628    &  34 &  \rm{m} &&5747 & 463 & 98 & \rm{m} \\ 

3271 & 1572    &  94 &  \rm{m} &&5835 &1519 & 94 & \rm{m} \\ 

3942 & 1552    &  98 &  \rm{m} &&5855 & 461 & 98 & \rm{m} \\ 

4150 & 1617    &  72 &  \rm{m} &&5887 & 569 & \phantom{0}1 & \rm{m}\\ 

4228 & 1461    &  98 &  \rm{m} &&5894 & 575 & 99 &\nodata \\ 

4292 &  803    &  98 &  \rm{m} &&5927 & 690 &  \phantom{0}0 & \rm{m} \\ 

4294 &  484    &  96 &  \rm{m} &&6175 & 359 &  \phantom{0}5 & \rm{m} \\ 

4346 &  691    &  98 &  \rm{m} &&6188 & 434 & 98 & \rm{m} \\ 

4524 & 1355    &  97 &  \rm{m} &&6353 & 362 & 89 & \rm{m} \\ 

4565 & 1353    &  96 &  \rm{m} &&6586 & 1223 & 95 & \rm{m} \\ 

4668 &  665    &  98 &  \rm{m} &&6602 & 1182 & 35 & \rm{m} \\

4671 &  672    &  98 &  \rm{m} &&6687 & 337 & \phantom{0}6 & \rm{m}\\ 

4705 &  643    &  98 &  \rm{m} &&6982 & 437 & 74 & \rm{m} \\ 

4756 &  473    &  71 &  \rm{m} &&8129 & 241 & 95 & \rm{m} \\ 

4829 & 1503    &  98 &  \rm{m} &&9159 &  73 & 88 & \rm{m} \\ 

4843 & 1488    &  75 &  \rm{m} &&9205 & 44  & 82 & \rm{m} \\ 

4909 & 1393    &  96 &  \rm{m} &&9291 &  35 & 53 &\nodata \\ 

5027 &  719    &  96 &  \rm{m} &&9351 &  88 & 78 & \rm{m} \\ 

5048 &  678    &  94 &  \rm{m} &&9401 &  20 & 73 & \rm{m} \\ 

5055 & 793 & 95 & \rm{m} && & & & \\

\enddata
\end{deluxetable}

\clearpage
\begin{deluxetable}{llll}
\tablewidth{6.0in}
\tablenum{7}
\pagestyle{empty}
\tablecaption{Basic astrometric data for NGC~188}
\tablehead{
\colhead{Parameter} & & \multicolumn {2} {c} {Value and its error }
  }
\startdata
Cluster center (J2000) & & $\alpha=0^{\rm h}47^{\rm m}12\fs5$ ($\pm3\fs6$)
\hspace{0.1in} & $\delta=+85\arcdeg14\arcmin49\arcsec$ ($\pm8\arcsec$)\\
Galactic coordinates  & & $l=122\fdg85$\hspace{0.1in} & $b=+22\fdg38$ \\
Absolute proper motion (mas~yr$^{-1}$) & & $\mu_\alpha\cos\delta=-2.56\pm0.2$
& $\mu_\delta=+0.18\pm0.2$ \\
No. of cluster members ($V<21$) & & \multicolumn {2} {c} {$\sim$1050} \\

\enddata
\end{deluxetable}

\end{document}